\documentclass[conference]{IEEEtran}
\IEEEoverridecommandlockouts
\usepackage{cite}
\usepackage{amsmath,amssymb,amsfonts}
\usepackage{algorithmic}
\usepackage{graphicx}
\usepackage{textcomp}
\usepackage{xcolor}

\usepackage{subfig}
\usepackage{multirow}

\newcommand{\rowname}[1]
{\rotatebox{90}{\makebox[.2\linewidth][c]{\textbf{#1}}}}
\newcommand\ainote[1]{\textcolor{red}{[AI: #1]}}
\newcommand\shnote[1]{\textcolor{blue}{[SH: #1]}}
\newcommand\kinnote[1]{\textcolor{green}{[KN: #1]}}
\newcommand{\ignore}[1]{}

\def\BibTeX{{\rm B\kern-.05em{\sc i\kern-.025em b}\kern-.08em
    T\kern-.1667em\lower.7ex\hbox{E}\kern-.125emX}}
\begin{document}

\title{Social-Media Activity Forecasting with Exogenous Information Signals\\

\thanks{Work supported by the DARPA SocialSim Program and the Air Force Research Laboratory under contract FA8650-18-C-7825.}
}

\author{\IEEEauthorblockN{Kin Wai NG}
\IEEEauthorblockA{\textit{Computer Science \& Eng.} \\
\textit{University of South Florida}\\
Tampa, USA \\
kinwaing@usf.edu}
\and
\IEEEauthorblockN{Sameera Horawalavithana}
\IEEEauthorblockA{\textit{Computer Science \& Eng.} \\
\textit{University of South Florida}\\
Tampa, USA \\
sameera1@usf.edu}
\and
\IEEEauthorblockN{Adriana Iamnitchi}
\IEEEauthorblockA{\textit{Computer Science \& Eng.} \\
\textit{University of South Florida}\\
Tampa, USA \\
anda@cse.usf.edu}
}

\maketitle

\begin{abstract}
Due to their widespread adoption, social media platforms present an ideal environment for studying and understanding social behavior, especially on information spread.
Modeling social media activity has numerous practical implications such as supporting efforts to analyze strategic information operations, designing intervention techniques to mitigate disinformation, or delivering critical information during disaster relief operations. 
In this paper we propose a modeling technique that forecasts topic-specific daily volume of social media activities by using both exogenous signals, such as news or armed conflicts records, and endogenous data from the social media platform we model. 
Empirical evaluations with real datasets from two different platforms and two different contexts each composed of multiple interrelated topics demonstrate the effectiveness of our solution.

\end{abstract}


\section{Introduction}
\ignore{
Social media platforms have become powerful communication channels in today’s society.
The ample volume of data available from these online platforms has offered an unparalleled opportunity to study social behavior, especially the interplay between real-world events and in-platform activity.
Researchers have widely used social media as open source indicators in various tasks, from predicting election outcomes~\cite{bermingham2011using} or fluctuations in the stock market~\cite{sul2017trading} to monitoring civil unrest and predicting protest and riot events~\cite{ramakrishnan2014beating}.
}
The large-scale distribution of information by social media platforms has offered a new public sphere for discussions and rapidly raising awareness of important issues, but at the same time exacerbated radical ideas and social divides.  
For example, the spread of disinformation aimed at increasing fragmentation and distrust between online communities~\cite{del2016spreading}, coordinated inauthentic efforts to manipulate the public such as cryptocurrency pump-and-dump online schemes~\cite{nizzoli2020charting}, casting doubts in electoral processes~\cite{ferrara2020characterizing}, or downplaying serious threats~\cite{tasnim2020impact}. 

Modeling social media activity has practical implications towards addressing challenges that currently overwhelm the digital space.
Forecasting how information spreads in the online world would not only offer more insights for understanding social behavior, but also enable risk control techniques to help flag suspicious activity or remove misleading content. 
For example, an accurate model that can predict how a particular topic will engage users and will generate activity in the near future can be used to flag unexpected spikes of user activity as potentially ignited by information operations aimed to divert attention or amplify emotions~\cite{beskow2020characterization}. 
Such modeling techniques can also be used for testing intervention techniques by social media platforms to safely experiment with questions such as which content promotion algorithms are inadvertently amplifying polarizing topics, which user accounts should be muted as they participate as unwitting crowds in a disinformation campaign, etc. 
Taking into account exogenous signals, such as political news or market indicators as well as signals from other social media platforms is paramount, given the tight interaction between multi-platform online activity and offline events.

This paper presents a topic-focused social media modeling solution that is part of the DARPA-sponsored SocialSim program~\cite{DARPA}.
We formulate the problem as a forecasting social media activity objective in order to evaluate how well it mirrors the real activity (the ground truth).
Our approach aims to forecast the daily volume of activity on a social media platform per given topic of discussion in a particular context. 
For example, the political crisis in Venezuela from the beginning of 2019 led to conversations on Twitter related to topics such as street protests, the validity of the presidential elections, and the impact of international aid in a country under economic and social duress. 
Our solution takes into account different sources of information exogenous to the social media platform whose activity we model, such as news or datasets of protests and armed conflicts. 
Moreover, from extensive data analysis~\cite{horawalavithanavz_websci21,dutta2020deep}, we are well aware that online topics are often stimulated by different exogenous signals over time, 
and no single exogenous source will be reliably mirrored in the activity of one particular topic on a social media platform. 
Our modeling approach reflects these data-driven observations. 

Our contributions are proposing and evaluating a modeling technique that forecasts topic-specific daily volume of activity for a longer period without relying on endogenous data during the prediction period. 
Thus, we forecast a week of activity at daily granularity using only the history of in-platform activity before the given week and contemporary exogenous signals. 
Via experimental evaluation on two real-world datasets, we show that our solution is generalizable to different social media platforms and to different contexts.
We demonstrate that while endogenous historical activity is indispensable for getting the right magnitude of activity, it is not sufficient for capturing occasional peaks of activity.
Thus, incorporating exogenous signals is critical.
Lastly, our experiments show that our solution fares better against proposed baselines in capturing both temporal scale and overall volume of activities.



\section{Related Work}
Previous studies developed many regression methods for the timeseries forecasting tasks~\cite{hewamalage2021recurrent}.
Popular statistical methods include Exponential Smoothing (ES) and the Autoregressive Integrated Moving Average (ARIMA). 
Unfortunately, these techniques suffer from limitations when handling non-linear forecasting problems due to the assumption that data are generated from linear processes~\cite{zhang1998forecasting}.

Several deep learning methods such as Convolutional Neural Networks (CNN) and Recurrent Neural Networks (RNN) have been proposed to tackle this problem for both univariate and multivariate timeseries prediction. 
Of those, Graph Convolutional Networks (GCNs), a generalized form of traditional CNNs, have attracted a lot of interest for univariate timeseries forecasting~\cite{li2017diffusion,yu2017spatio}.
Li et al.~\cite{li2017diffusion} proposed a GCN architecture called Diffusion Convolutional Recurrent Neural Networks (DCRNN) and showed state-of-the-art performance in traffic forecasting tasks.
The authors claimed that such GCN model could be generalized to any univariate forecasting tasks with graph data available.   
However, Hernandez et al.~\cite{hernandez2020using} applied this technique on multiple social media datasets, and showed its poor performance of forecasting user activity timeseries.
They show that the performance degrades depending on the heterogeneity of user activity.

Agent-based-modeling (ABM) techniques use both statistical and machine learning regression methods to forecast individual user activity streams.
Abdelzaher et al.~\cite{abdelzaher2020multiscale} represent each user's activity by a timeseries of K elements, where each element represents the user activity in an arbitrary time granularity (e.g., hours, days, etc.).
They used both ARIMA and deep neural networks (such as CNN and RNN) to predict the next K elements of the timeseries.
Moreover, separate models were implemented to capture the activity streams of different users.
This approach did not scale well when there are millions of users who participate in social media discussions.

Other studies predict the popularity of topics~\cite{liu2015effectively,yin2013unified}, hashtags~\cite{kong2014predicting}, or keywords~\cite{saleiro2016learning} shared on Twitter messages.
Liu et al.~\cite{liu2015effectively} explore several machine learning methods to predict whether and when a topic will become prevalent.
The authors highlight the challenges faced on forecasting the frequency of topics discussed by users due to irregular patterns.
Yin et al.~\cite{yin2013unified} demonstrate that topics prevalent on Twitter can be categorized into temporal topics (e.g., breaking events) and stable topics (e.g., user interests).
They utilize both the network structure and temporal information to predict whether a topic is temporal or stable. 
Given the classification nature of their task, this model is not directly applicable in activity forecasting settings.
However, their analysis reveals interesting properties of topics that could be leveraged as features when forecasting topic-specific activities.
Saleiro et al.~\cite{saleiro2016learning} classify the popularity of named entity mentions (e.g., "Ronaldo") on Twitter as high or low in the following hours using the features extracted from news articles.
They found news carry different predictive power based on the nature of the entities under study.
Dutta et al.~\cite{dutta2020deep} predict the volume of Reddit discussions in a future short-time horizon leveraging the text from news and initial set of comments using a recurrent neural network architecture.
Shrestha et al.~\cite{shrestha2019learning} used a deep learning model to forecast the number of retweets and mentions of a specific news source on Twitter using the network structure observed in the day before the predictions.
They found that small, but dense network structures are helpful in the predictions.

Although these machine learning approaches have shown promising results, they have been exclusively applied to binary classification tasks (e.g., predicting topic popularity) or short-term forecasting (e.g., predicting the final size of information cascades, or forecasting next day of activity assuming immediate past information is always available).
This work investigates forecasting long-term social media activity at the topic level.


\section{Problem Definition and Modeling}

Our goal is to forecast social media activity on a given platform as it pertains to various topics of discussion.
Specifically, we plan to forecast the daily volume of activities per topic as would happen in different social media platforms and across different contexts.
We choose to use machine learning in order to provide an easily generalizable solution across platforms and social phenomena.

The machine learning task is thus the following: given the history of daily activity streams on a given topic and platform, as well as a set of exogenous features, predict a sequence of future daily events within a given time period

\subsection{Model Design}
\label{sec:design}

As shown in previous work~\cite{horawalavithanavz_websci21,dutta2020deep}, online topics of discussion are often correlated with different exogenous signals over time.
For example, in the context of the Venezuela political crisis, Horawalavithana et al.~\cite{horawalavithanavz_websci21} show that Twitter discussions related to the announcement of Juan Guaid\'{o} as interim president were dictated by particular external events and news articles reports while Twitter discussions related to Venezuela's former president, Ch\'{a}vez, were strongly correlated with Reddit activity.  
One single trained machine learning model cannot capture these subtle relationships between different topics and exogenous/endogenous sources.
Our objective is to dynamically choose between different sources of information to predict topic activity.
To this end,
we proposed TAP (Topic Activity Predictors), a design that consists of a collection of temporal neural network models trained with different subsets of exogenous/endogenous sources. 
We chose LSTMs~\cite{hochreiter1997long} as our machine learning architecture due to its capability of learning temporal relationships from multivariate data, handling multi-step predictions, and testing different lagged observations as input time steps.

\begin{table}[htpb]
\centering
\caption{Feature set used to train TAP models. All time series are in daily granularity.}
\label{tab:features}
\begin{tabular}{|l|l|}
\hline
\textbf{Source} & \textbf{Feature Description} \\ \hline
\multirow{2}{*}{\textbf{\begin{tabular}[c]{@{}l@{}}News/\\ GDELT\end{tabular}}} & \begin{tabular}[c]{@{}l@{}}The time series for each GDELT event \\ (20 in total) as recorded in~\cite{schrodt2012cameo}.\end{tabular} \\ \cline{2-2} 
 & News articles volume time series for a given topic. \\ \hline
\multirow{2}{*}{\textbf{Reddit}} & Posts volume time series for a given topic. \\ \cline{2-2} 
 & Comments volume time series for a given topic. \\ \hline
\multirow{2}{*}{\textbf{ACLED}} & \begin{tabular}[c]{@{}l@{}}The volume time series of events grouped by scale\\  (i.e., local, regional, national, and international).\end{tabular} \\ \cline{2-2} 
 & \begin{tabular}[c]{@{}l@{}}The volume time series for each ACLED event \\ (6 in total) as recorded in~\cite{raleigh2010introducing}.\end{tabular} \\ \hline
\multirow{2}{*}{\textbf{\begin{tabular}[c]{@{}l@{}}Social Media\\ Platforms\end{tabular}}} & New user volume time series for a given topic. \\ \cline{2-2} 
 & Shares volume time series for a given topic. \\ \hline
\end{tabular}
\end{table}

Our solution uses timeseries from both real-world events and social media platform activities as features to forecast the number of online daily activities for a given topic in the future. 
Moreover, a one-hot encoding vector was used as a static input feature in all models to represent different topics of discussion. 
Additionally, we trained one LSTM model for each exogenous/endogenous source.
The list of exogenous and endogenous features is shown  in Table~\ref{tab:features}.

An important design consideration in our solution was to include models that favor the learning of both short-term and long-term time dependencies. 
To this end, we consider models with different lookback factors defined by $m$, specifically 14, 7 and 3, as well as different multistep-ahead prediction windows defined by $n$, specifically 7, 3 and 1.
Thus, each model will generate a forecast for a given interval of $n$ days based on a sequence of historical daily observations of length $m$.
Given that we do not have access to the target variable (ground truth) during the testing period, the models' predictions are fed back to our solution until the last time step to forecast is reached. 

The final number of trained models per social media platform in our pool is given by the product of the total number of exogenous sources available and the number of time windows combinations in consideration.
A critical step in our solution is to dynamically select which model from the pool is best suitable to generate forecasts for a given topic.
This decision is based on the models' performance over a held-out validation set, which was set to one week prior to the forecasting window.
We choose RMSE as the error metric to compare between models and consequently select the one that provides the lowest error on validation. 
In our experiments, RMSE produced better results for model selection than other error metrics tested (e.g., MSE, SMAPE, etc.).
However, the choice of the right error metric might differ based on the problem or different datasets.
Thus, the error metric selection represents an additional parameter to consider in our solution.

\ignore{
\begin{figure}[htbp]
    \centering
    \includegraphics[width=0.98\linewidth]{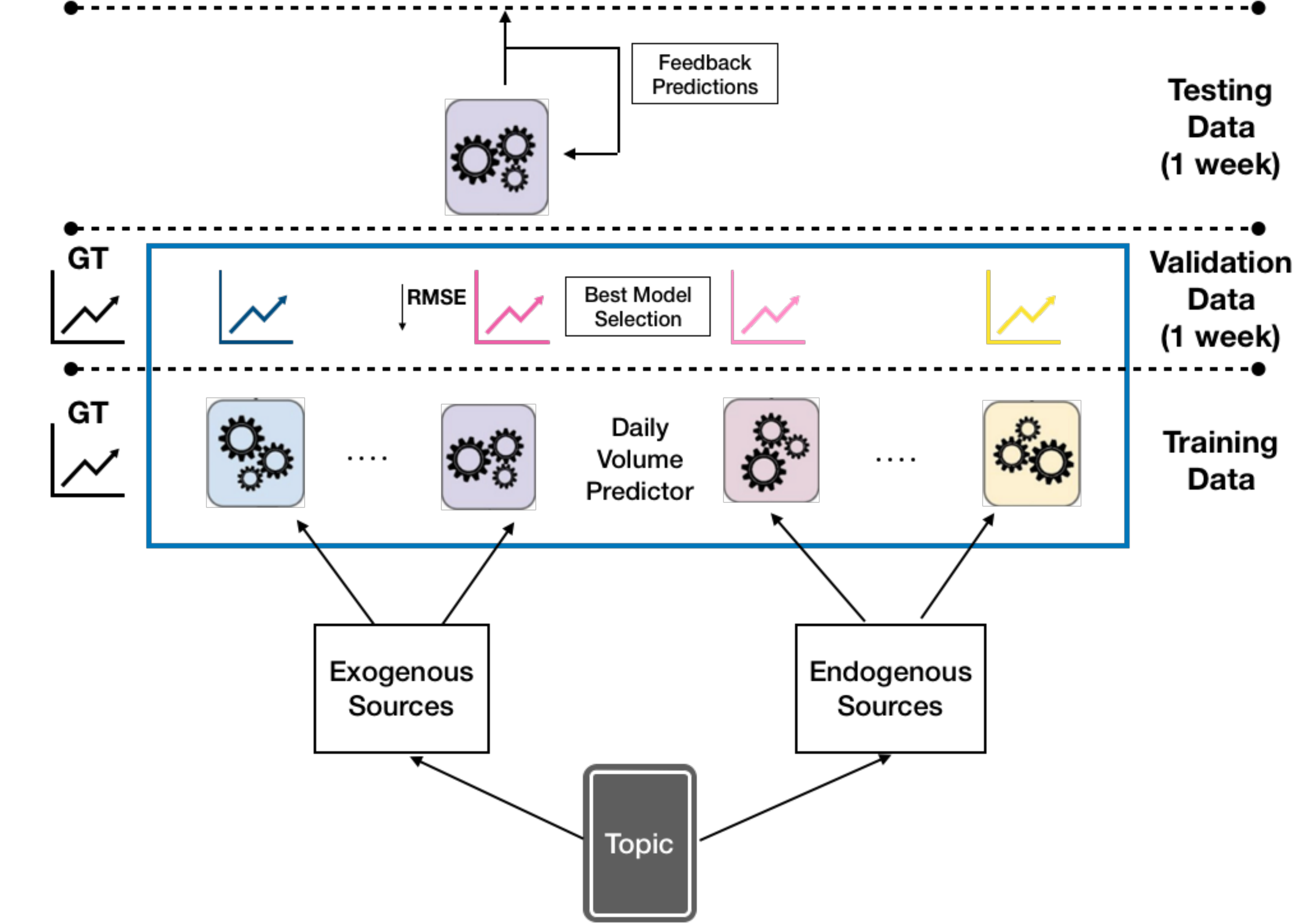}
    \caption{Selecting the best model to forecast the daily volume of discussions per topic. We develop a variety of forecasting models depending on the exogenous/endogenous sources and time window for predictions. For each topic, we select the best model depending on its performance in the validation period.}
    \label{fig:mcas_model_selection}
\end{figure}
}

\begin{figure}[htbp]
    \centering
    \includegraphics[width=0.98\linewidth]{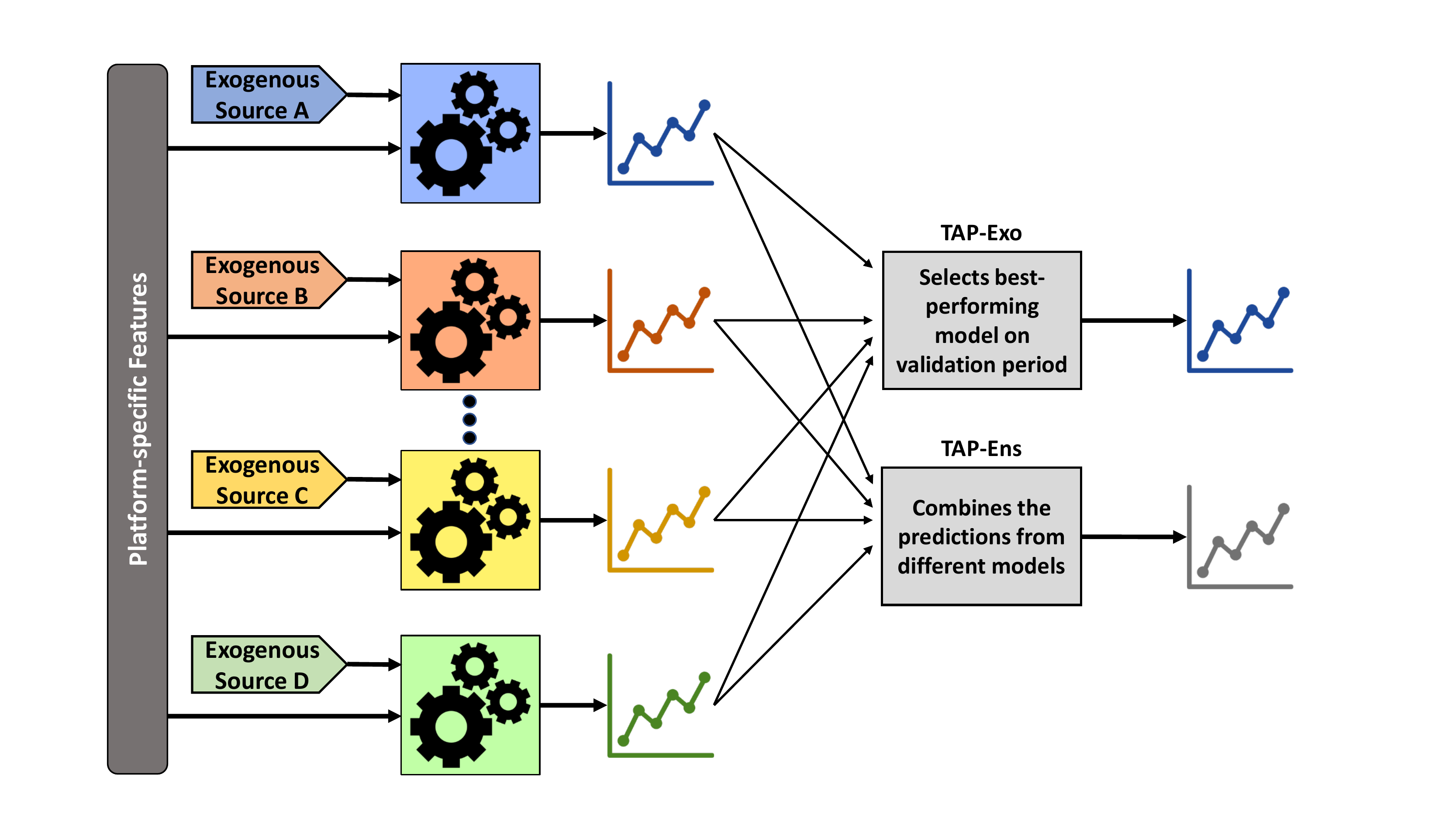}
    \caption{Model diagrams for TAP-Exo and TAP-Ens.}
    \label{fig:mcas_model_diagram}
\end{figure}

Our objective is to evaluate the effect of exogenous and endogenous features in forecasting social media activity. 
Thus, we implemented three different design variants that attempt to solve the problem from different perspectives, such as only looking at in-platform activity traces, leveraging a specific contemporary exogenous signal, or combining a number of exogenous sources together. The model designs are illustrated in Figure~\ref{fig:mcas_model_diagram}.
These models are as following:
\begin{itemize}
    \item \textbf{TAP-Exo} dynamically chooses an exogenous source and a particular prediction horizon to forecast the activity for a given topic. The selection is made based on the performance on a validation set. The intuition is that different topics are stimulated by different exogenous signals over time. 
    \item \textbf{TAP-Endo} uses only internal features corresponding to a specific social media platform. The best internal model for a given topic and a particular time window is chosen based on the performance on a validation set. The intuition is that historical social media activity for a given topic is enough to generate accurate predictions.   
    \item \textbf{TAP-Ens} is an ensemble of models that takes the average over the predictions produced by our collection of independent exogenous models. The intuition is that combining the predictions from models leveraging different exogenous sources as features is more important for accurate predictions. 
\end{itemize}

\section{Datasets}

We use data from two different geo-political contexts, each with their own different behavior on social media.
First, we focus on the 2019 Venezuela Political Crisis (VZ).
For the last two decades, the Venezuelan society has experienced a pervasive sociopolitical fragmentation fueled by differences of interests, identities, and politics.
In Venezuela, the political spectrum is for the most part divided into two parties: Chavism -- those who support the political ideology of the late president Hugo Chavez, and Anti-Chavism -- those strongly opposed to Chavez's legacy.
Today Chavism still maintains control of the Venezuelan political system with Nicolas Maduro as the head of state.
However, failure to manage globalization, lack of investment in infrastructure, and a poor administration has put the country in the grip of a significant economic collapse.
As a result, it has contributed to unprecedented conflicts and high political tension which resulted in nationwide protests, militarized responses, and incidents of mass violence and arrests in early 2019.
During this period, social media platforms were used as a critical medium for mass mobilization by political parties and leaders.
For Venezuelans, the digital space became a valuable source of information to stay informed and specially communicate the daily events developing in the country.

Second, we use data related to the China-Pakistan Economic Corridor (CPEC).
The CPEC project seeks to transform Pakistan's economy by modernizing its road, air and energy transportation systems, and at the same time expand China's trading power through connecting the deep-sea Pakistani ports to China's province.
Heavily criticized due to its opaque nature~\cite{afzal2020all,hameed2018politics},
conflict around this project plays out in conversations on social media, where both Pakistani and Chinese state actors have worked together to control and promote the political narrative around CPEC~\cite{david_2021}.

From these contexts, we use data from social media and also exogenous sources that record real-time events. 

\subsection{Twitter and YouTube Data Collection}

\begin{figure}[!t]
    \subfloat[Venezuela]{
        \includegraphics[width=0.48\columnwidth]{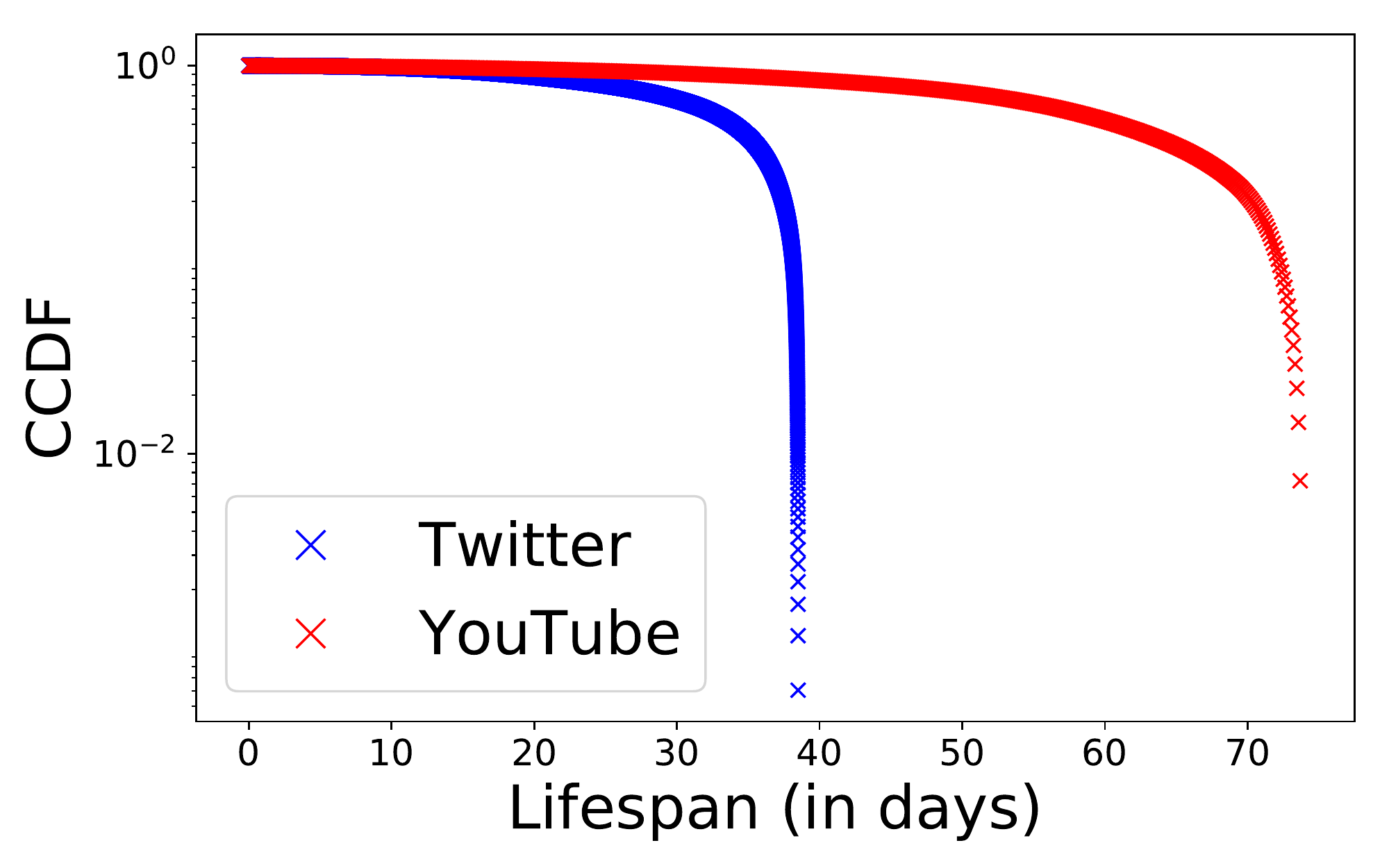}
        \label{fig:vz_lifespan}
    }
    \subfloat[CPEC]{
        \includegraphics[width=.48\columnwidth]{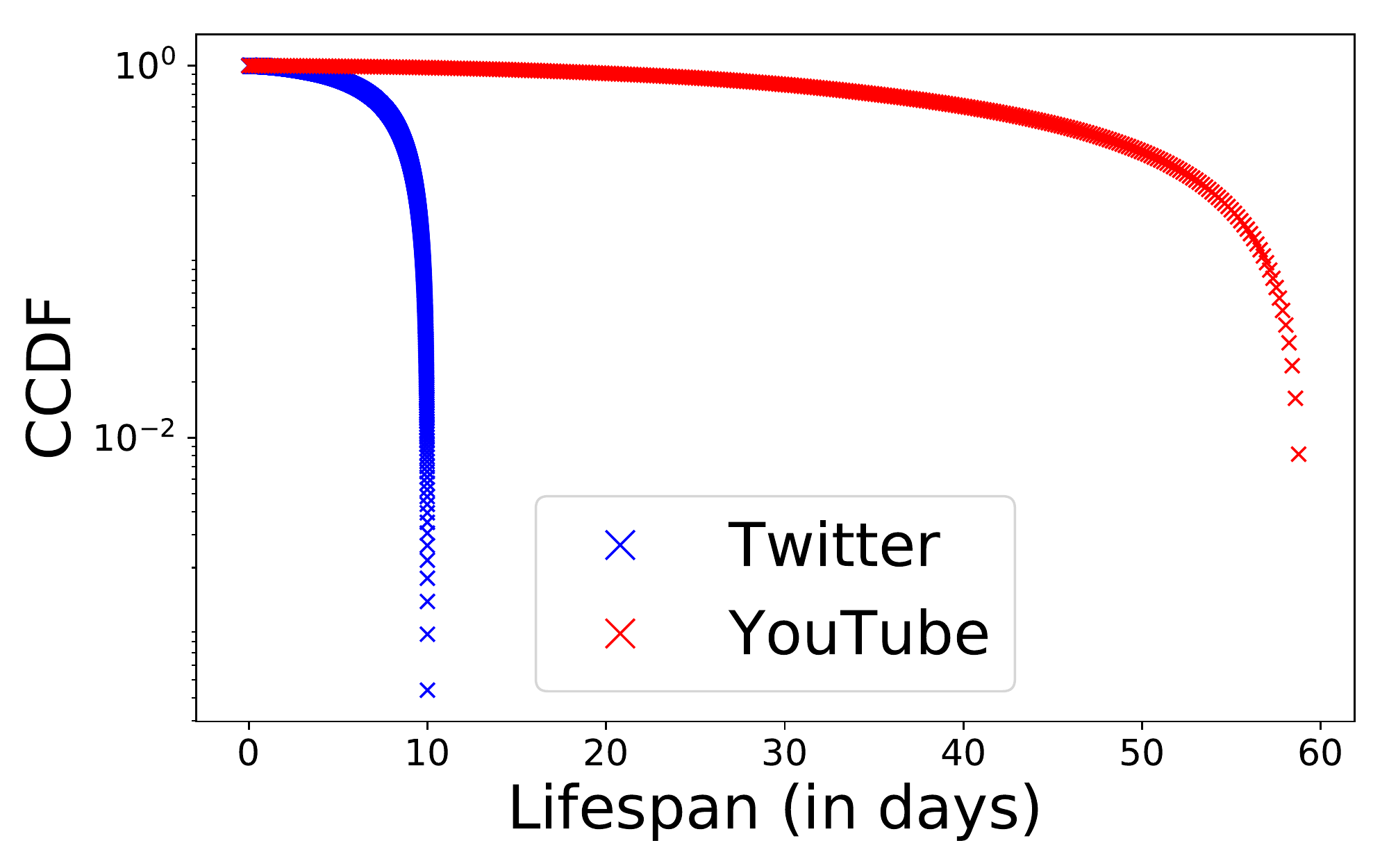}
        \label{fig:cpec_lifespan}
    }
    \caption{Lifespan distribution of tweets and videos for VZ and CPEC datasets.}
    \label{fig:lifespan_ccdf}
\end{figure}

Twitter data was collected using the GNIP API based on a list of relevant keywords compiled by subject matter experts (SME) across the two contexts of interest.
YouTube data was collected by using its public API query tool for a list of keywords generated by SME.
Both Twitter and YouTube datasets are completely anonymized to protect users privacy.
For the VZ dataset, the collection period was from December 24th, 2018 to March 7th, 2019.
This dataset consists of 20,562,479 Twitter activities (5.4\% tweets and 94.6\% responses including retweets, quotes and replies) done by 1,077,398 users, and a total of 165,298 YouTube activities (1.4\% videos and 98.6\% comments) done by 77,356 users.
The CPEC dataset was collected from March 30th, 2020 to August 3rd, 2020 and it consists of 5,386,114 Twitter activities (4.4\% tweets 95.6\% responses) done by 1,221,365 users, and 303,214 YouTube activities (1.1\% videos and 98.9\% comments) done by 149,374 users.
We note that both Twitter and YouTube are platforms with very different characteristics, specially when it comes to their internal temporal logic. 
Figure~\ref{fig:lifespan_ccdf} presents the complementary cumulative distribution of tweets and videos regarding their lifespan across two different contexts. 
We observe that tweets exhibit a much shorter lifespan (a median of 1 day in both contexts) than YouTube videos (a median of 9 days in VZ and 19 days in CPEC).
While posts in Twitter quickly fade away, videos in YouTube often act as an archive having a much longer life.
These differences allow to investigate the generalizability of our approach not only on different contexts, but also on social media platforms with contrasting behaviors.

To identify the most representative topics originating from social media discussions specific to VZ and CPEC, a thorough exploration of the datasets corpus was conducted by SME. 
For VZ, a total of 18 topics were identified as follows: \emph{protests}, \emph{military}, \emph{violence}, \emph{guaido/legitimate}, \emph{other/censorship\_outage}, \emph{other/chavez}, \emph{arrests}, \emph{international/aid}, \emph{international/respect\_sovereignty}, \emph{arrests/opposition}, \emph{maduro/legitimate}, \emph{other/chavez/anti}, \emph{military/desertions}, \emph{other/anti\_socialism}, \emph{maduro/narco}, \emph{international/aid\_rejected}, \emph{maduro/cuba\_support}, and \emph{maduro/dictator}.
In CPEC, the annotation effort resulted in 12 topics: \emph{benefits/development/energy}, \emph{benefits/development/roads}, \emph{benefits/jobs}, \emph{controversies/china/border}, \emph{controversies/china/uighur}, \emph{controversies/pakistan/bajwa}, \emph{controversies/pakistan/baloch}, \emph{controversies/pakistan/students}, \emph{leadership/bajwa}, \emph{leadership/sharif}, \emph{opposition/propaganda}, and \emph{other}.
 
Because it is not feasible to manually label millions of messages, a semi-supervised classification task was conducted consisting of two steps: (1) manually annotating an initial subset of messages, and (2) training a multilingual BERT model to classify each message with one or multiple topics for each dataset.
The manual annotation process was conducted over a corpus of 11,218 messages in each dataset, and consisted in a 8 to 1 ratio of single-annotator annotations to all-annotator annotations. 
The 18 topics identified in VZ dataset reported inter-annotator agreement scores of 0.64 for the weighted average Cohen’s Kappa, and 0.7 for the Fleiss’Kappa measurement.
The 12 topics in CPEC reported inter-annotator agreement scores of 0.75 for the weighted average Cohen’s Kappa.
Previous work has also found similar Cohen's Kappa agreement scores in a variety of datasets on sentiment analysis annotation tasks  \cite{ribeiro2016sentibench}.
After manual annotation, a BERT model was trained for topic annotation~\cite{devlin2019bert}. 
The BERT model was trained on 9,535 distinct text documents and evaluated on a 15\% test set (1,682 texts) in each dataset. 
Stratified sampling was used to ensure that the train and test sets have approximately the same percentage of samples of each topic class as in the original manually annotated corpus. 
The model obtained a precision of 67\%, recall of 66\%, and F1 score of 66\% in the VZ dataset, and a precision of 80\%, recall of 68\%, and F1 score of 73\% in the CPEC dataset.
Tables~\ref{tab:vz_dataset_distribution}~and~\ref{tab:cpec_dataset_distribution} show the distribution of activities across topics for Vz and CPEC, respectively.

We emphasize that this effort of labeling messages with semantically-representative topics is used in our approach only to ensure a meaningful grouping of messages under topics. 
Our solution does not use semantic information, thus being generalizable to totally different topics and contexts.

\begin{table*}[htpb]
\centering
\caption{Distribution of activities across topics in the Venezuela dataset.}
\label{tab:vz_dataset_distribution}
\begin{tabular}{lrrrr}
\hline
\multicolumn{1}{c}{\multirow{2}{*}{Topic}} & \multicolumn{2}{c}{Counts} & \multicolumn{2}{c}{Frequency} \\
\multicolumn{1}{c}{} & \multicolumn{1}{c}{Twitter} & \multicolumn{1}{c}{YouTube} & \multicolumn{1}{c}{Twitter} & \multicolumn{1}{c}{YouTube} \\ \hline
military & 3,629,731 & 13,789 & 17.6\% & 8.3\% \\
protests & 2,839,647 & 4,119 & 13.8\% & 2.5\% \\
international/aid & 2,442,336 & 13,279 & 11.9\% & 8\% \\
violence & 2,416,929 & 13,774 & 11.8\% & 8.3\% \\
guaido/legitimate & 1,857,828 & 14,357 & 9\% & 8,7\% \\
maduro/dictator & 1,343,551 & 28,255 & 6.5\% & 17.1\% \\
international/respect\_sovereignty & 1,324,265 & 3,932 & 6.4\% & 2.4\% \\
other/chavez & 1,027,070 & 21,515 & 5\% & 13\% \\
arrests & 866,012 & 2,145 & 4.2\% & 1.3\% \\
arrests/opposition & 672,937 & 672 & 3.3\% & 0.4\% \\
maduro/legitimate & 573,368 & 478 & 2.8\% & 0.3\% \\
international/aid\_rejected & 383,368 & 3,094 & 1.9\% & 1.9\% \\
other/chavez/anti & 344,479 & 14,568 & 1.7\% & 8.8\% \\
maduro/narco & 218,665 & 4,587 & 1.1\% & 2.8\% \\
other/anti\_socialism & 206,780 & 24,125 & 1\% & 14.6\% \\
military/desertions & 195,044 & 502 & 0.9\% & 0.3\% \\
maduro/cuba\_support & 113,392 & 1,938 & 0.6\% & 1.2\% \\
other/censorship\_outage & 107,077 & 169 & 0.5\% & 0.1\% \\ \hline
\end{tabular}
\end{table*}

\begin{table*}[htpb]
\centering
\caption{Distribution of activities across topics in the CPEC dataset.}
\label{tab:cpec_dataset_distribution}
\begin{tabular}{lrrrr}
\hline
\multicolumn{1}{c}{\multirow{2}{*}{Topic}} & \multicolumn{2}{c}{Counts} & \multicolumn{2}{c}{Frequency} \\
\multicolumn{1}{c}{} & \multicolumn{1}{c}{Twitter} & \multicolumn{1}{c}{YouTube} & \multicolumn{1}{c}{Twitter} & \multicolumn{1}{c}{YouTube} \\ \hline
other & 2,793,092 & 299,319 & 51.9\% & 98.7\% \\
controversies/china/border & 1,436,463 & 928 & 26.7\% & 0.3\% \\
controversies/pakistan/baloch & 210,315 & 708 & 3.9\% & 0.2\% \\
opposition/propaganda & 198,270 & 262 & 3.7\% & 0.1\% \\
leadership/sharif & 154,309 & 468 & 2.9\% & 0.2\% \\
benefits/development/roads & 148,971 & 667 & 2.8\% & 0.2\% \\
controversies/china/uighur & 140,183 & 213 & 2.6\% & 0.1\% \\
benefits/development/energy & 131,941 & 345 & 2.4\% & 0.1\% \\
benefits/jobs & 80,652 & 85 & 1.5\% & 0.03\% \\
leadership/bajwa & 49,562 & 190 & 0.9\% & 0.06\% \\
controversies/pakistan/bajwa & 30,130 & 24 & 0.6\% & 0.01\% \\
controversies/pakistan/students & 12,226 & 5 & 0.2\% & 0.0\% \\ \hline
\end{tabular}
\end{table*}

\subsection{Exogenous Sources}

\begin{table}[htpb]
\centering
\caption{Size of exogenous datasets.}
\label{tab:exogenous_datasets}
\begin{tabular}{|l|r|r|}
\hline
\multicolumn{1}{|c|}{\multirow{2}{*}{Dataset}} & \multicolumn{2}{c|}{Counts} \\ \cline{2-3} 
\multicolumn{1}{|c|}{} & \multicolumn{1}{c|}{VZ} & \multicolumn{1}{c|}{CPEC} \\ \hline \hline
News Articles & 141,197 & 168,351 \\ \hline
ACLED & 803 & 2,977 \\ \hline
Reddit posts & 5,598 & 26,744 \\ \hline
Reddit comments & 59,092 & 628,159  \\ \hline

\end{tabular}
\end{table}

As exogenous information, we used two sources that reflect real-world events, news and armed conflicts, and another social media platform, Reddit. Table~\ref{tab:exogenous_datasets} shows the number of activities collected in each source.

The news articles data was collected via a publicly available geopolitical event database, GDELT~\cite{leetaru2013gdelt}. 
The GDELT database was queried for events containing either "Venezuela" or "Pakistan" keywords, where the event occurred during the same time period as our social media datasets.
For each article, we also have the source text and the date when it was published.


Armed Conflict Location and Event Dataset (ACLED) records violent (e.g., protests, riots, political repressions) and non-violent events (e.g., strategic developments) carried out by political agents, including government officials, rebels, and militias~\cite{raleigh2015armed}.
These reports are collected primarily from local and regional news sources, Integrated Regional Information Network, Relief Web, Factiva, and humanitarian agencies.

Reddit is a popular website focused on content creation and threaded discussions. 
Content is organized in topic-specific bulletin boards called subreddits.
For CPEC, data was collected by searching submissions and comments containing geo-political keywords related to CPEC. 
Additionally, the \texttt{/r/pakistan} subreddit was collected in its entirety for the collection period.
For VZ, we collected discussions structured around one of the largest Venezuela-related subreddits, \texttt{/r/vzla}.

We assigned topics to exogenous data by running the previously pre-trained BERT model over the content of news articles and Reddit messages.
This step enables the extraction of  information  signals  relevant  to  a  particular  topic  thus  reducing potential noise from the exogenous data.
We randomly sampled 500 Reddit posts and news articles to check the reliability of this classification in each context.
We were only able to find 5\% false positives.
\section{Evaluation}


Our solution is based on a collection of 12 LSTM models per social media platform, from which 9 leverage endogenous and selected exogenous sources while the remaining 3 use only endogenous historical platform activity. 

\subsection{Experimental Setup}

For all LSTM models, we used the same space of hyper-parameters.
We trained for 200 epochs with a starting learning rate of $1e^{-3}$.
The optimizer used in the framework was Adam and the loss function used was MAE (Mean Absolute Error). 
Each model consisted of one LSTM layer followed by a fully connected layer to generate the predictions.
We tested multiple numbers of hidden units (30, 10 and 5), and the best was chosen based on this validation set.
For models trained with exogenous sources, we assume that previous day exogenous data information is always available.
For both datasets, we split the time series samples into a training and testing period.
In the Venezuela dataset, the train period is from 2018-12-24 to 2019-02-14 (53 days), and the test period is from 2019-02-15 to 2019-03-07 (21 days).
In the CPEC dataset, the train period is from 2020-03-30 to 2020-07-13 (106 days), and the test period is from 2020-07-14 to 2020-08-03 (21 days).
The week prior to the start of each forecasting period was used as a validation set.
The models were set to forecast 1-week of social media activity, and this process was repeated for the 3 consecutive weeks in the testing period.

We report three performance metrics in the results tables.
Specifically, we measure the absolute percentage error (APE), which evaluates  the error over the volume of activities aggregated over a particular time period, root mean squared error (RMSE), which takes into account the error in predictions over time, and symmetric mean absolute percentage error (SMAPE), which also takes into account relative errors in the temporal scale.
These metrics were calculated over each 1-week interval in the testing period for each topic and then averaged across the three weeks of predictions.

\subsection{Baselines}
We used the following three baselines for comparison.

\emph{Persistent Baseline} repeats 7-days of platform activity. 
It assumes that the activity observed over a period of one week will also remain the same in the following week.

\emph{Autoregressive Integrated Moving Average (ARIMA)} is a popular  statistical model commonly applied to forecast the price or return of stocks. 
This model considers past observations when making future predictions.
We tuned the parameters of this model using our validation set.
Particularly, we ran a grid search over $p$ \{0-7\} (the number of lag observations), $d$ \{0-2\} (the degree of differencing), and $q$ \{0-2\} (the size of the moving average window).

\emph{Hawkes Processes} have been applied in a variety of settings in order to describe or predict  univariate or multivariate data. It was originally defined to describe earthquake dynamics~\cite{hawkes1971spectra}, but previous work has used it for forecasting tasks such as estimating price fluctuations in stocks~\cite{bacry2015hawkes} or reproducing conversation dynamics as event sequences~\cite{masuda2013self}.

\subsection{Forecasting Daily Activities on Twitter and YouTube}

We report experimental results for forecasting the number of activities across two different online platforms in Table~\ref{tab:shares_performance} for both VZ and CPEC contexts.
The table shows the average model performance over various topics in terms of three forecasting metrics.
On Twitter, we observe that our models outperform all baselines across all metrics in consideration.  

Hawkes processes are the most competitive method in Twitter out of the three baselines considered.
However, there is a notable performance gain by each of our models when compared to this baseline. 
For instance, the best performing model in VZ (TAP-Ens) achieves a 28\% and 16\% improvement over the baseline for APE and RMSE metrics, respectively. 
Moreover, in CPEC, the best performing model (TAP-Exo) shows a 59\% improvement over Hawkes on APE for capturing the overall volume of activities.
Nevertheless, Hawkes remains competitive in capturing temporal patterns as TAP-Exo only shows an improvement of about 1\% on RMSE. 
On YouTube, however, our models' performance on predicting the overall volume of activities was not significantly better than the best performing baseline in each context.
Only the TAP-Ens model achieved a comparable accuracy to baselines on APE while TAP-Exo achieved similar performance as Hawkes in CPEC.
TAP-Ens and TAP-Endo seem to fare well against baselines on the temporal scale, as they obtained the lowest RMSE and SMAPE scores.

\begin{table}[!t]
\centering
\caption{
Mean performance across various topics of interest in three different forecasting metrics. The performance on a given topic was averaged across the three weeks that were predicted. The best-3 performing models are bold.}
\label{tab:shares_performance}
\begin{tabular}{|c|c|r|r|r|r|}
\hline
\textbf{Domain} & \textbf{Model} & \multicolumn{1}{c|}{\textbf{APE}} & \multicolumn{1}{c|}{\textbf{SMAPE}} & \multicolumn{1}{c|}{\textbf{RMSE}} \\ \hline
 & TAP-Exo & \textbf{110.58} & \textbf{91.56} & \textbf{26331.52} \\
 & TAP-Endo & \textbf{107.26} & \textbf{88.26} & \textbf{24206.13}  \\
VZ & TAP-Ens & \textbf{89.88} & \textbf{78.68} & \textbf{22472.26}  \\
Twitter & ARIMA & 302.76 & 101.63 & 28108.81 \\
 & Hawkes & 124.76 & 93.31 & 26911.77 \\
 & Persistent & 399.04 & 100.94 & 44245.76 \\ \hline
 & TAP-Exo & 92.96 & \textbf{80.92} & 203.82  \\
 & TAP-Endo & \textbf{63.25} & \textbf{69.11} & \textbf{171.74} \\
VZ & TAP-Ens & \textbf{58.76} & \textbf{66.75} & \textbf{165.57}  \\
YouTube & ARIMA & 79.61 & 90.26 & 198.36 \\
 & Hawkes & 80.36 & 119.86 & 241.87 \\
 & Persistent & \textbf{51.73} & 89.42 & \textbf{186.65}  \\ \hline
 & TAP-Exo & \textbf{76.26} & \textbf{102.91} & \textbf{3065.56} \\
 & TAP-Endo & \textbf{102.38} & \textbf{99.62} & \textbf{2560.24} \\
CPEC & TAP-Ens & \textbf{146.27} & \textbf{101.04} & \textbf{2950.47}  \\
Twitter & ARIMA & 236.93 & 115.21 & 4121.61 \\
 & Hawkes & 187.58 & 103.99 & 3084.78 \\
 & Persistent & 825.25 & 121.18 & 10096.03 \\ \hline
 & TAP-Exo & \textbf{51.38} & 83.61 & 122.62 \\
 & TAP-Endo & \textbf{53.16} & \textbf{69.54} & \textbf{114.10} \\
CPEC & TAP-Ens & \textbf{50.13} & \textbf{78.65} & \textbf{116.03} \\
YouTube & ARIMA & 87.12 & \textbf{81.65} & \textbf{121.54} \\
 & Hawkes & \textbf{51.38} & 95.58 & 157.65 \\
 & Persistent & 66.55 & 95.00 & 143.66 \\ \hline
\end{tabular}
\end{table}

In some cases, large prediction errors caused by a few topics might skew the overall mean performance of a model.
Thus, we also evaluate models performance based on their average ranking across topics.
Particularly, we ranked the models according to their performance for each topic.
The best model is assigned the rank of 1, the second best rank 2, and so on.
Lastly, we compute the average rankings across all topics for each platform and context.
Table~\ref{tab:friedman_ranking_performance} shows that TAP-Ens model is the most consistent as it is ranked 1st in 5 out of 8 possible cases, and it is always among the top-3 best models.
Next up is the TAP-Endo model which is frequently among the top-3 models, and it is followed by TAP-Exo. 
Overall, these results suggest that our models tend to fare better than baselines in both temporal trend and the total volume of activities across topics. 

\begin{table}[htpb]
\centering
\caption{The average ranking scores for each model across topics. Lower ranking scores are better.}
\label{tab:friedman_ranking_performance}
\begin{tabular}{|c|cc|cc|}
\hline
\textbf{Domain} & \textbf{Rank \textsc{(rmse)}} & \textbf{Model} & \textbf{Rank \textsc{(ape)}} & \textbf{Model} \\ \hline
\multirow{6}{*}{\begin{tabular}[c]{@{}c@{}}VZ\\ Twitter\end{tabular}} & 1.61 & TAP-Ens & 2.33 & TAP-Ens \\
 & 3.0 & TAP-Endo & 3.0 & TAP-Exo \\
 & 3.38 & Hawkes & 3.11 & TAP-Endo \\
 & 3.55 & TAP-Exo & 3.61 & Hawkes \\
 & 3.94 & ARIMA & 3.77 & ARIMA \\
 & 5.5 & Persistent & 5.16 & Persistent \\ \hline
\multirow{6}{*}{\begin{tabular}[c]{@{}c@{}}VZ\\ YouTube\end{tabular}} & 2.27 & TAP-Ens & 2.66 & Persistent \\
 & 2.77 & TAP-Endo & 2.94 & TAP-Ens \\
 & 3.5 & Persistent & 3.0 & TAP-Endo \\
 & 3.55 & ARIMA & 3.94 & Hawkes \\
 & 4.0 & TAP-Exo & 4.22 & TAP-Exo \\
 & 4.88 & Hawkes & 4.22 & ARIMA \\ \hline
\multirow{6}{*}{\begin{tabular}[c]{@{}c@{}}CPEC\\ Twitter\end{tabular}} & 2.33 & TAP-Endo & 2.33 & TAP-Exo \\
 & 2.41 & TAP-Ens & 2.83 & TAP-Ens \\
 & 2.75 & TAP-Exo & 2.83 & TAP-Endo \\
 & 4.0 & Hawkes & 3.66 & Hawkes \\
 & 4.16 & ARIMA & 4.58 & ARIMA \\
 & 5.33 & Persistent & 4.75 & Persistent \\ \hline
\multirow{6}{*}{\begin{tabular}[c]{@{}c@{}}CPEC\\ YouTube\end{tabular}} & 2.0 & TAP-Ens & 2.9 & TAP-Ens \\
 & 2.7 & TAP-Endo & 3.2 & Persistent \\
 & 3.6 & ARIMA & 3.4 & TAP-Exo \\
 & 3.9 & TAP-Exo & 3.7 & TAP-Endo \\
 & 4.2 & Hawkes & 3.9 & ARIMA \\
 & 4.6 & Persistent & 3.9 & Hawkes \\ \hline
\end{tabular}
\end{table}

\begin{figure*}[htpb]
\centering
\begin{tabular}{ccc}
	\subfloat[international/aid (VZ Twitter)]{
		\includegraphics[width=0.28\linewidth]{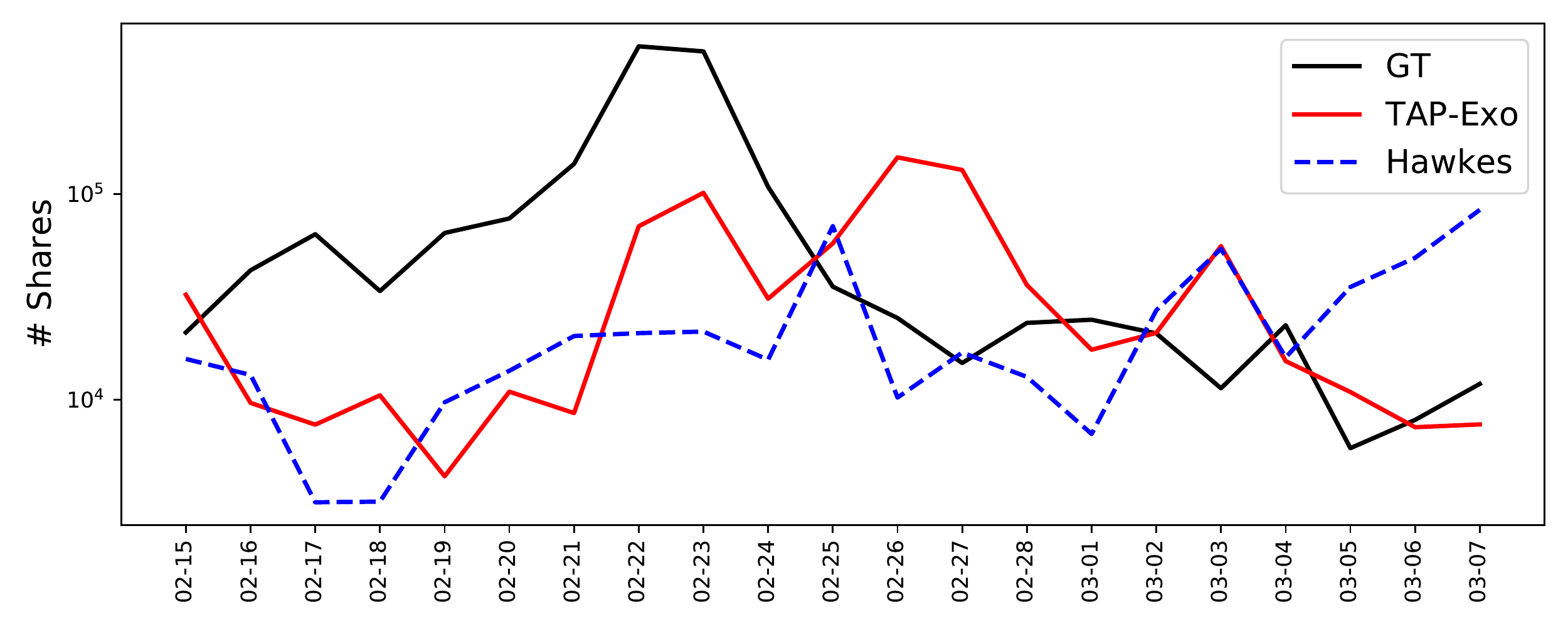}
			\label{fig:international-aid_twitter_vz}
	}
	&
\subfloat[other/chavez (VZ Twitter)]{
		\includegraphics[width=0.28\linewidth]{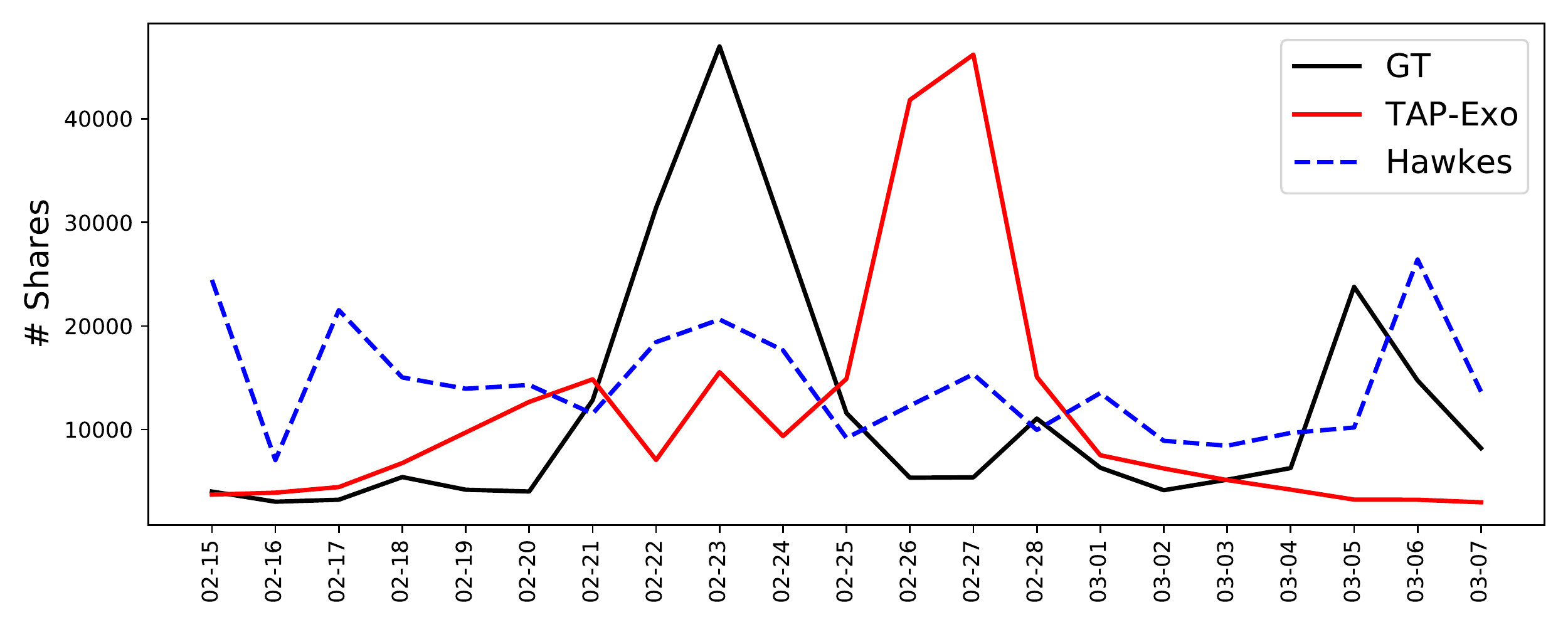}
			\label{fig:other-chavez_twitter_vz}
	}
	&
	\subfloat[military/desertions (VZ YouTube)]{
		\includegraphics[width=0.28\linewidth]{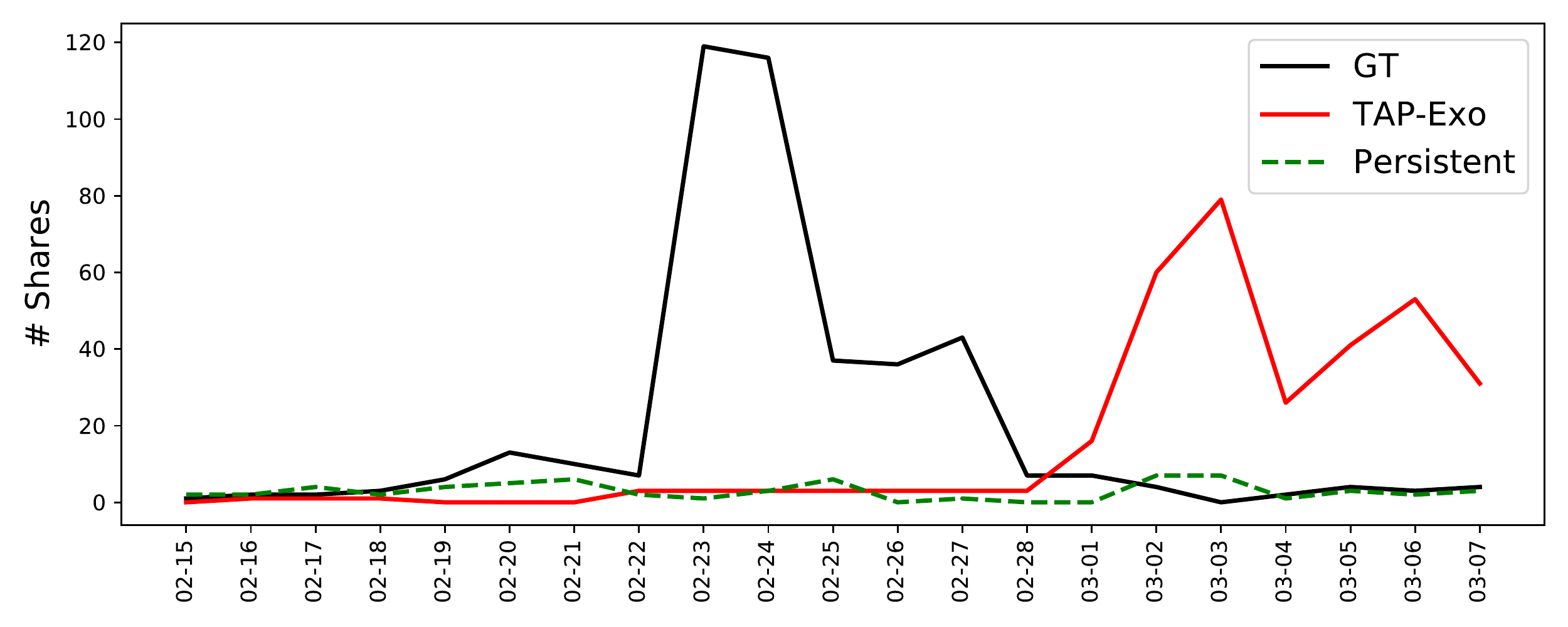}
			\label{fig:military-desertions_yt_vz}
	}
\\
\subfloat[controversies/pakistan/baloch (CPEC Twitter)]{
		\includegraphics[width=0.28\linewidth]{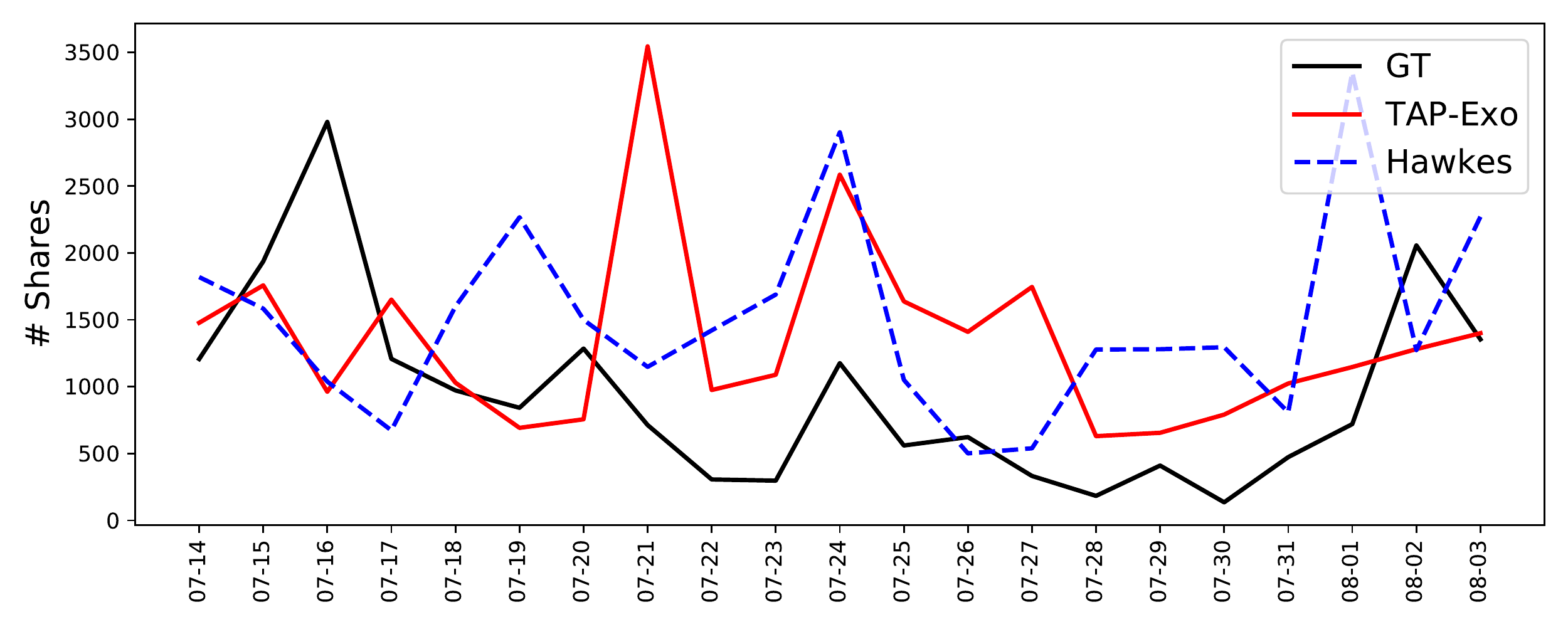}
			\label{fig:controversies-pakistan-baloch_twitter_cpec}
	}
	&
	\subfloat[benefits/development/roads (CPEC YouTube)]{
		\includegraphics[width=0.28\linewidth]{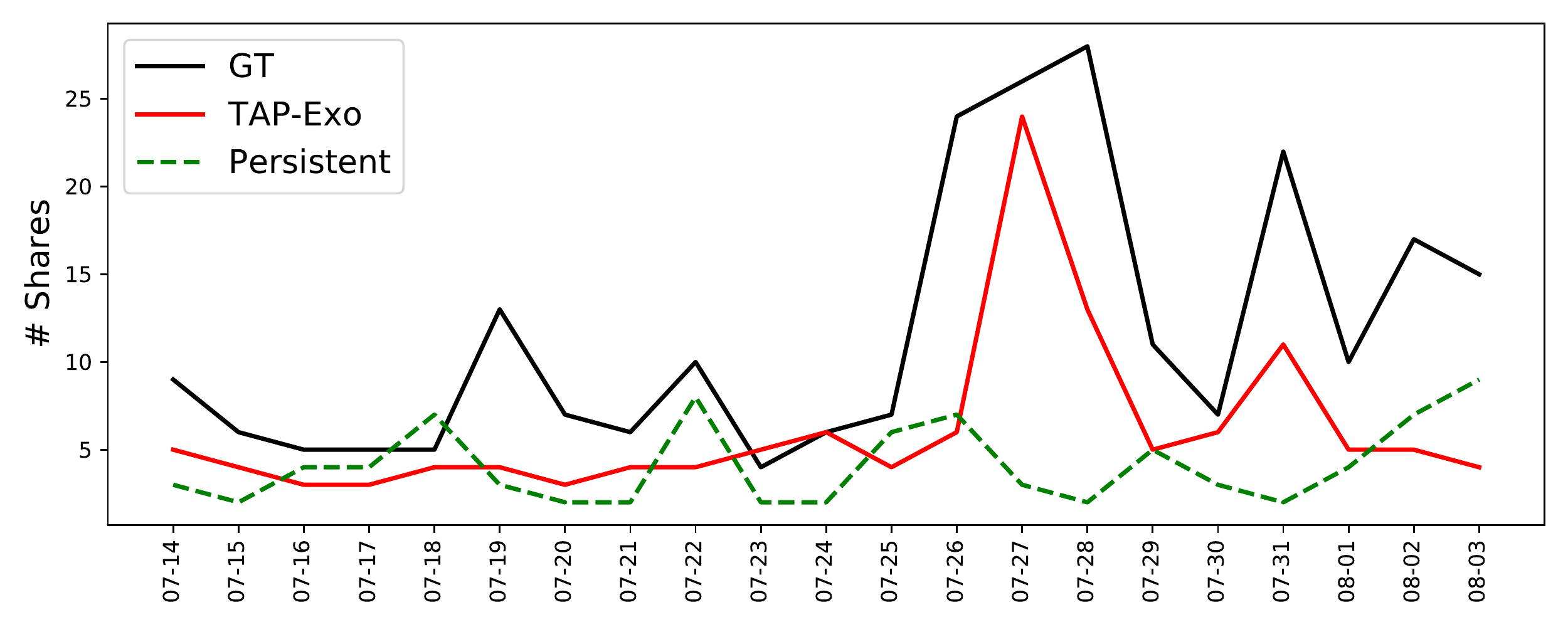}
			\label{fig:benefits-development-roads_yt_cpec}
	}
	&
\subfloat[leadership/sharif (CPEC YouTube)]{
		\includegraphics[width=0.28\linewidth]{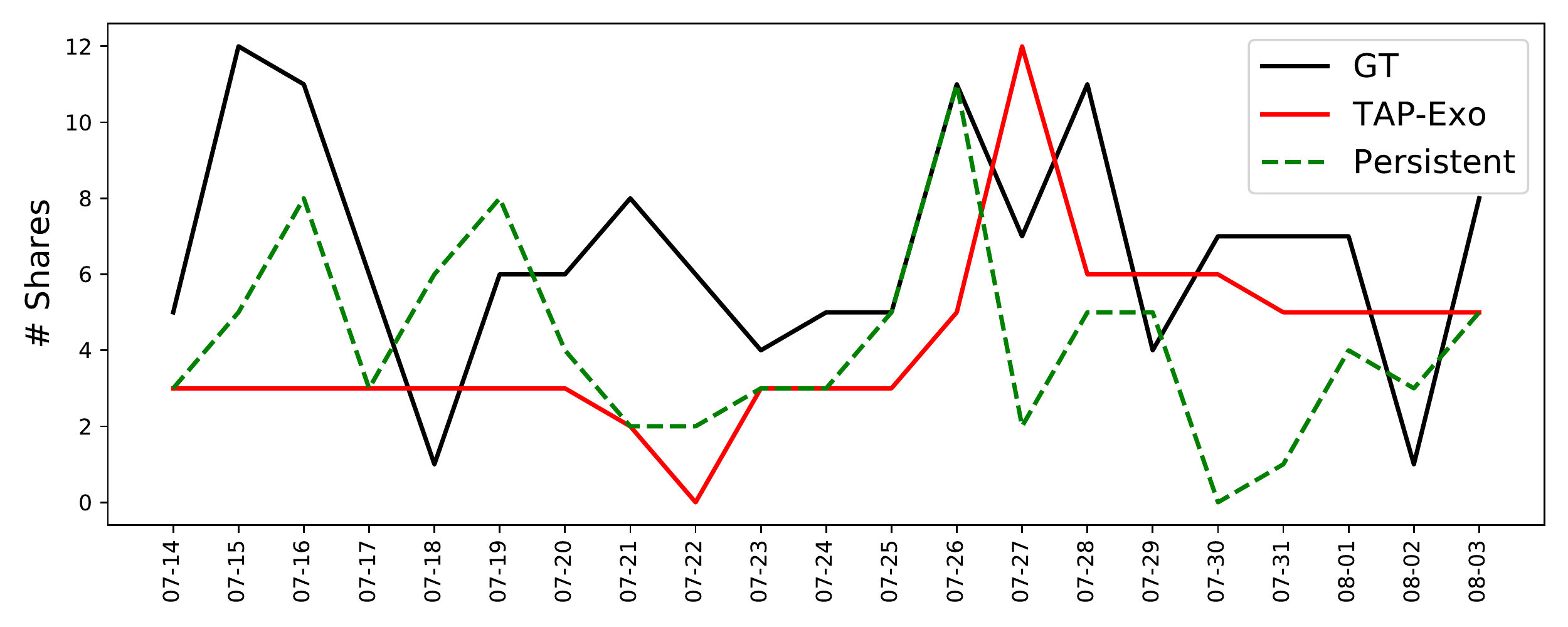}
			\label{fig:leadership-sharif_yt_cpec}
	}
\\
\end{tabular}
	 \caption{Sample time series visualizations of Twitter and YouTube predictions for number of shares across two different contexts. Black line - ground truth, red line - TAP predictions, green line - persistent baseline activity, and blue line - Hawkes predictions.}
	 \label{fig:sample_timeseries}
\end{figure*}

\subsection{Topic-Level Evaluation Performance}

\begin{table*}[htpb]
\centering
\caption{Per topic RMSE performance on number of shares. 
Cells are bold for the top-3 best performing models.}
\label{tab:shares_rmse_topic_performance}
\begin{tabular}{lcccccc}
\hline
\multicolumn{1}{|l}{} & \multicolumn{1}{l}{} & \multicolumn{3}{c}{\textbf{Venezuela Twitter}} & \multicolumn{1}{l}{} & \multicolumn{1}{l|}{} \\ \hline
\multicolumn{1}{|l|}{\textbf{Model}} & \textbf{military} & \textbf{international/aid} & \textbf{\begin{tabular}[c]{@{}c@{}}maduro/\\ cuba\_support\end{tabular}} & \textbf{guaido/legitimate} & \textbf{other/chavez} & \multicolumn{1}{c|}{\textbf{arrests/opposition}} \\ \hline
\multicolumn{1}{|l|}{TAP-Exo} & \textbf{82475.95} & \textbf{109673.50} & \textbf{4191.62} & 30527.90 & 13327.20 & \multicolumn{1}{c|}{4493.59} \\
\multicolumn{1}{|l|}{TAP-Endo} & \textbf{69851.68} & \textbf{100245.30} & \textbf{4030.62} & 24008.93 & 11989.04 & \multicolumn{1}{c|}{\textbf{4208.44}} \\
\multicolumn{1}{|l|}{TAP-Ens} & \textbf{71387.17} & \textbf{105253.20} & \textbf{3827.76} & \textbf{22053.58} & \textbf{10208.37} & \multicolumn{1}{c|}{\textbf{3150.65}} \\
\multicolumn{1}{|l|}{ARIMA} & 90227.49 & 131401.43 & 6348.49 & \textbf{21652.94} & \textbf{9076.91} & \multicolumn{1}{c|}{\textbf{2592.57}} \\
\multicolumn{1}{|l|}{Hawkes} & 88491.17 & 120778.54 & 4379.73 & \textbf{19641.50} & \textbf{11062.74} & \multicolumn{1}{c|}{9919.60} \\
\multicolumn{1}{|l|}{Persistent} & 104836.70 & 118525.82 & 4565.25 & 61052.77 & 47580.38 & \multicolumn{1}{c|}{27841.27} \\ \hline
\multicolumn{1}{|l}{} & \multicolumn{1}{l}{} & \multicolumn{3}{c}{\textbf{Venezuela YouTube}} & \multicolumn{1}{l}{} & \multicolumn{1}{l|}{} \\ \hline
\multicolumn{1}{|l|}{\textbf{Model}} & \textbf{protests} & \textbf{international/aid} & \textbf{maduro/narco} & \textbf{\begin{tabular}[c]{@{}c@{}}international/\\ aid\_rejected\end{tabular}} & \textbf{military} & \multicolumn{1}{c|}{\textbf{military/desertions}} \\ \hline
\multicolumn{1}{|l|}{TAP-Exo} & \textbf{42.99} & \textbf{424.55} & 113.23 & \textbf{137.85} & 443.34 & \multicolumn{1}{c|}{39.32} \\
\multicolumn{1}{|l|}{TAP-Endo} & \textbf{35.70} & 531.16 & \textbf{98.32} & \textbf{136.74} & \textbf{359.17} & \multicolumn{1}{c|}{26.93} \\
\multicolumn{1}{|l|}{TAP-Ens} & \textbf{42.19} & \textbf{443.58} & \textbf{95.71} & 143.16 & \textbf{335.96} & \multicolumn{1}{c|}{27.49} \\
\multicolumn{1}{|l|}{ARIMA} & 53.37 & \textbf{473.48} & 131.45 & \textbf{129.17} & 360.19 & \multicolumn{1}{c|}{\textbf{23.35}} \\
\multicolumn{1}{|l|}{Hawkes} & 89.56 & 529.08 & 140.96 & 153.34 & 503.42 & \multicolumn{1}{c|}{\textbf{25.91}} \\
\multicolumn{1}{|l|}{Persistent} & 113.50 & 492.80 & \textbf{108.77} & 156.67 & \textbf{300.69} & \multicolumn{1}{c|}{\textbf{24.63}} \\ \hline
\multicolumn{1}{|l}{} & \multicolumn{1}{l}{} & \multicolumn{3}{c}{\textbf{CPEC Twitter}} & \multicolumn{1}{l}{} & \multicolumn{1}{l|}{} \\ \hline
\multicolumn{1}{|l|}{\textbf{Model}} & \textbf{\begin{tabular}[c]{@{}c@{}}controversies/\\ pakistan/baloch\end{tabular}} & \textbf{\begin{tabular}[c]{@{}c@{}}opposition/\\ propaganda\end{tabular}} & \textbf{\begin{tabular}[c]{@{}c@{}}controversies/\\ china/uighur\end{tabular}} & \textbf{\begin{tabular}[c]{@{}c@{}}benefits/develop/\\ roads\end{tabular}} & \textbf{leadership/sharif} & \multicolumn{1}{c|}{\textbf{\begin{tabular}[c]{@{}c@{}}controversies/\\ china/border\end{tabular}}} \\ \hline
\multicolumn{1}{|l|}{TAP-Exo} & \textbf{926.60} & \textbf{495.29} & \textbf{3898.53} & 2828.81 & 1954.54 & \multicolumn{1}{c|}{\textbf{1460.22}} \\
\multicolumn{1}{|l|}{TAP-Endo} & \textbf{976.81} & \textbf{525.22} & \textbf{4267.74} & \textbf{2653.42} & \textbf{1442.12} & \multicolumn{1}{c|}{\textbf{2819.94}} \\
\multicolumn{1}{|l|}{TAP-Ens} & \textbf{935.46} & \textbf{541.76} & \textbf{4108.38} & 2711.62 & \textbf{1588.20} & \multicolumn{1}{c|}{5791.20} \\
\multicolumn{1}{|l|}{ARIMA} & 1042.04 & 646.05 & 5343.69 & \textbf{2709.76} & \textbf{1519.02} & \multicolumn{1}{c|}{\textbf{1595.91}} \\
\multicolumn{1}{|l|}{Hawkes} & 1091.68 & 1284.49 & 4338.48 & \textbf{2461.50} & 1761.78 & \multicolumn{1}{c|}{5328.16} \\
\multicolumn{1}{|l|}{Persistent} & 5807.37 & 3901.98 & 5043.89 & 2878.01 & 3189.86 & \multicolumn{1}{c|}{63761.06} \\ \hline
\multicolumn{1}{|l}{} & \multicolumn{1}{l}{} & \multicolumn{3}{c}{\textbf{CPEC YouTube}} & \multicolumn{1}{l}{} & \multicolumn{1}{l|}{} \\ \hline
\multicolumn{1}{|l|}{\textbf{Model}} & \textbf{benefits/jobs} & \textbf{leadership/bajwa} & \textbf{\begin{tabular}[c]{@{}c@{}}benefits/develop/\\ roads\end{tabular}} & \textbf{\begin{tabular}[c]{@{}c@{}}opposition/\\ propaganda\end{tabular}} & \textbf{leadership/sharif} & \multicolumn{1}{c|}{\textbf{\begin{tabular}[c]{@{}c@{}}controversies/\\ china/uighur\end{tabular}}} \\ \hline
\multicolumn{1}{|l|}{TAP-Exo} & \textbf{1.24} & \textbf{3.33} & \textbf{7.13} & 2.18 & 4.19 & \multicolumn{1}{c|}{6.62} \\
\multicolumn{1}{|l|}{TAP-Endo} & \textbf{1.08} & \textbf{3.35} & 9.25 & 2.10 & 4.08 & \multicolumn{1}{c|}{6.53} \\
\multicolumn{1}{|l|}{TAP-Ens} & \textbf{1.19} & \textbf{3.41} & \textbf{8.51} & \textbf{1.78} & \textbf{3.87} & \multicolumn{1}{c|}{\textbf{6.49}} \\
\multicolumn{1}{|l|}{ARIMA} & 1.32 & 12.42 & 24.70 & \textbf{1.82} & \textbf{3.01} & \multicolumn{1}{c|}{\textbf{5.83}} \\
\multicolumn{1}{|l|}{Hawkes} & 1.36 & 4.29 & \textbf{8.70} & \textbf{2.06} & 5.04 & \multicolumn{1}{c|}{\textbf{6.43}} \\
\multicolumn{1}{|l|}{Persistent} & 1.92 & 5.81 & 9.94 & 2.94 & \textbf{3.94} & \multicolumn{1}{c|}{6.72} \\ \hline
\end{tabular}
\end{table*}

Even though our models perform well against baselines, the results tend to vary across different topics. 
We investigate this more closely by looking at topic-wise model performance in terms of RMSE. 
Table~\ref{tab:shares_rmse_topic_performance} reports the results for our models and baselines across various topics.
We present results for the best-3 and worst-3 performing topics in each context and social media platform. 
Additionally, we present time series visualizations for some of these topics in Figure~\ref{fig:sample_timeseries}.
We include only the predictions from the TAP-Exo model to highlight the importance of contemporary exogenous data for predicting spikes of activity.
For comparison, we also include the  time  series  from  the  overall best  performing  baseline across  each  social  media  platform, specifically Hawkes (for Twitter) and the persistent baseline (for YouTube).  
From this table and figures, we have numerous observations.

In the context of Venezuela, we found that \emph{military}, \emph{international/aid}, and \emph{maduro/cuba\_support} were among the best predicted topics on Twitter.
TAP-Endo achieved consistently the lowest RMSE scores out of all models and baselines across these topics.
This suggests that only using historical information from the online platform itself might be enough to obtain good performance.
We explored TAP-Endo predictions more closely by looking at \emph{international/aid}, one of the most active topics of Twitter discussions in Venezuela during our testing period.
Particularly, this topic spanned a few days of exceptionally large spikes of activity that were correlated with specific real-life events developing in the country.
We found that TAP-Endo was not capable of capturing these large and rapid bursts of activity, but instead it predicted well the relative periods of low-activity present in the topic.  
On the other hand, TAP-Exo was more accurate on approximating the temporal pattern exhibited in this topic as shown in Figure~\ref{fig:international-aid_twitter_vz}.
We believe this is due to TAP-Exo's choice of leveraging news articles and ACLED events as features which greatly helped with predicting these rare spikes of events.
Nevertheless, TAP-Endo still managed to rank higher than TAP-Exo with respect to RMSE performance, which highlights the need for investigating different evaluation metrics that could look at different levels of success. 

For the worst performing topics, we found that only 4 out of 18 had one baseline outperforming all of our models. 
The worst predicted topics in VZ were \emph{guaido/legitimate}, \emph{other/chavez}, and \emph{arrest/opposition}.
We visually inspected our models' predictions on these topics to gain insights about their performance. 
In some cases we found that our models' predictions seem to match the actual ground truth better than the best performing baseline.  
For example, TAP-Exo was the only model able to predict that a large spike of activity would happen for \emph{other/chavez} topic during the forecasting interval (as shown in Figure~\ref{fig:other-chavez_twitter_vz}).
We noticed that our TAP-Exo model dynamically chose Reddit to predict this topic across all three weeks.
From data analysis, we found that Reddit activities and Twitter discussions related to \emph{other/chavez} were strongly correlated during the training period.
Hence, our model's decision to use only Reddit and ignore other exogenous sources helped greatly with predicting spikes of activity on this topic as would happen in the future.
Nevertheless, TAP-Exo was heavily penalized due to being off in its predictions by a few days, while baselines such as Hawkes and ARIMA obtained lower RMSE scores despite predicting nearly a flat line throughout the forecasting interval. 

In the context of CPEC,~\emph{controversies/pakistan/baloch},~\emph{opposition/propaganda} and~\emph{controversies/china/uighur} are the best-performing topics on Twitter.
While the two controversies topics are related to human rights abuse against Balochistan people in Pakistan, and Uyghurs in Xinjiang, China, the other topic is related to discussions that promote the negative coverage of CPEC characterizing it as propaganda and misinformation.
On these three topics, TAP-Exo showed the best performance overall.
Reddit features were deemed by our model the most useful for both \emph{controversies/china/uighur} and \emph{opposition/propaganda} topics, and this observation remained constant across the three weeks of predictions.
On the other hand, the exogenous sources selected for \emph{controversies/pakistan/baloch} varied across the 1-week prediction intervals, specifically ACLED, News, and Reddit were chosen for the 1st, 2nd and 3rd week, respectively.
This dynamic selection of different exogenous signals helped with approximating the actual temporal pattern observed in this topic (as shown in Figure~\ref{fig:controversies-pakistan-baloch_twitter_cpec}). 
Out of the 12 topics in CPEC, there is only one instance (i.e., \emph{benefits/development/roads}) in which a baseline managed to rank first on Twitter.
However, the performance differences between our models and this baseline were not significantly large.

On YouTube, TAP-Exo beats at least one baseline in 16 out of 18 topics in VZ, and 9 out of 12 topics in CPEC, in terms of RMSE.
We noticed that TAP-Exo model was the best among other model variants, specially compared to the model that used only endogenous features.
As shown in Figures~\ref{fig:benefits-development-roads_yt_cpec}~and~\ref{fig:leadership-sharif_yt_cpec}, we observe that TAP-Exo tends to capture relatively well both the temporal trend and scale on two active topics \emph{benefits/development/roads} and~\emph{leadership/sharif}.
These topics covered discussions about infrastructure projects in Pakistan related to CPEC, and the leadership by Nawaz Sharif, former prime minister of Pakistan, who initiated the CPEC project in 2015.
Our model picked both Reddit and News articles to predict~\emph{leadership/sharif} topic, and ACLED to predict~\emph{benefits/development/roads}. 

The persistent baseline was the most competitive as it repeats the recent past for many topics.
Our models performed poorly on some less active topics due to sparse activity. 
For example, one of the  worst performing topics in Venezuela is \emph{military/desertions}.
We note that the forecasting performance was relatively close across models and baselines. 
Figure~\ref{fig:military-desertions_yt_vz} shows the actual time series for \emph{military/desertions} during the testing period.
In general, all models, with the exception of TAP-Exo, predicted a time series similar to the one exhibited by the baseline.
For this topic, the TAP-Exo model considered news and GDELT features to be more useful for predictions.
TAP-Exo's predictions were penalized the most due to overpredicting the third week of the testing period.

\section{Summary and Discussion}

We presented our approach for modeling per-topic activity on social media platforms developed as part of the DARPA-funded SocialSim research program. 
Our approach is generalizable to different social media platforms; it is generalizable to different contexts; and it adapts to dynamically changing exogenous influences. 
We demonstrated these properties by evaluating our solution for forecasting the daily volume of user activities on two different social media platforms with different response rates and types of engagement: Twitter and YouTube. 
Moreover, we used user activities from two different contexts, one created by a political crisis, the other by political efforts to promote international economic cooperation in some Asian countries. 
As exogenous signals we used publicly available datasets of related news and armed conflicts, in addition to Reddit activities on corresponding subreddits. 

Our experimental evaluations show the following:
While endogenous historical activity is indispensable for getting the right magnitude for daily activity, it is not sufficient for forecasting peaks of activity. 
For this, the right exogenous signal can help. 
We also showed that our solutions that use exogenous information outperformed the model that only uses endogenous information in 91 out of 120 cases.
Moreover, in most cases our models outperformed the baselines.

With a simulator that accurately forecasts social media activity per topic of discussion, we plan as future work to develop techniques that will allow reliable testing of interventions in social media platforms with the scope of containing disinformation and other information operations. 

\section{Acknowledgments}
This work is supported by the DARPA SocialSim Program and the Air Force Research Laboratory under contract FA8650-18-C-7825. The authors would like to thank Leidos for providing data.

\bibliographystyle{IEEEtran}
\bibliography{bib-v1.bib}

\end{document}